# Algebraic Determination of Spectral Characteristics of Rovibrational States of Diatomic Molecules. I. Diagram Technique for Determination of Vibrational Dependences of Matrix Elements

## S. A. Astashkevich[*]

St. Petersburg State University, Peterhof, St. Petersburg, 198904 Russia

## Abstract

Explicit algebraic expressions for the expansion of the vibrational matrix elements $\langle v|f(r)|v'\rangle$ in series of matrix elements on the wave functions of the ground vibrational state have been obtained for arbitrary sufficiently differentiable functions of the internuclear distance $f(r)$, arbitrary values $v$ and $v'$, and the potential curves whose ladder operators can be constructed. A diagram technique have been developed for it that consists in: 1) the numeration of the matrix elements

$$M_{l,k+m}^{k-l} = \langle l \left| \frac{d^{k-l}}{dy^{k-l}} f(r(y)) \right| k+m \rangle$$

($y$ - a dynamic variable) by points of the 2D diagram with coordinates $(l, k)$, 2) the drawing arrows between points of this diagram corresponding to the action of the annihilation operators $\hat{K}^-$ on the wave functions; 3) total taking into account of all possible path vectors formed by the continuous sequences of arrows from point $(v, v)$ towards points $(0, k)$. The only requirement is that the action of the operator $\hat{K}^-$ on the wave functions should give the wave functions of the Schrödinger equation with the potential curve having the same parameter values. All necessary data for algebraic calculations of the vibrational dependence of matrix elements for the harmonic oscillator and the Morse potential have been given. Obtained expressions can be used to determine the absolute values and vibrational dependences of various spectroscopic characteristics of both ground and electronically excited states of diatomic molecules.

**Keywords**: vibrational matrix elements, algebraic methods, ladder operators, diagram technique.

PACS number(s): 03.65.Fd, 03.65.Ca, 33.20.-t

---

[*] *E-mail address:* astashkevich@mail.ru

# 1. Introduction

Along with the using of methods based on the numerical solution of the vibrational Schrödinger equation [1, 2] for the calculation of matrix elements corresponding to different characteristics of molecules algebraic methods are actively developed in last time [3–5]. These methods include the factorization [5, 6], group theory [7], the super-symmetric quantum mechanics (SUSY) [8] and other methods. The basic point of these methods is to find analytical solutions of the Schrödinger equation for the energy levels, the wave functions and matrix elements using the formalism of the quantum mechanics and mathematical physics (including the hypervirial theorems, the Hellmann-Feynman theorem, the sum rules formulas and etc.; the second quantization method, analytical properties of special (hypergeometric) functions, the theory of Lie algebra operators, etc.). It is important to emphasize that the algebraic methods, in contrast to numerical methods, provide a theoretical possibility of finding an explicit dependences of the matrix elements on vibrational and rotational quantum numbers. In particular this makes the algebraic methods are promising for the development of semi-empirical methods of determination of vibrational and rotational dependences of the spectroscopic characteristics of molecules. It is also important to note that the algebraic methods, in principle, allow developing an error estimations mathematical device for calculations of matrix elements using information about the inaccuracy of used potential energy curve parameters and some functions corresponding to matrix elements on the electronic wave functions of a certain quantum-mechanical operators (dependences on internuclear distance of the dipole moment, the transition dipole moment and others).

It is important to note that algebraic methods can be effectively applied to describe not only the energy of molecular levels but also non-energetic (radiative, electrical, magnetic) characteristics of the vibrational and vibrational-rotational levels (see bibliography in monographs [3–5]). This is especially important taking into account that the non-energetic characteristics, in particular, the radiative transitions probabilities and the g-factors of molecular levels may have significantly higher information content about the internuclear dynamics than the energetic characteristics [9, 10]. Our cycle of papers devoted in the first place to the algebraic description namely non-energetic characteristics of the vibrational, vibrational-rotational and rovibrational states of diatomic molecules.

It should be noted that the published algebraic expressions for the vibrational dependence of matrix elements of an arbitrary function of have been obtained only in the case of the harmonic potential [11–13]. For anharmonic potentials algebraic expressions for dependences of the vibrational matrix elements on vibrational [14−22] and rotational [23−28] quantum numbers of the combining states were obtained only for some model functions (polynomial, exponential, and some others) for the following potentials: quartic anharmonic oscillator [14], Morse [15−20, 23−26], modified Pöschl−Teller [21, 22]), and Kratzer [27, 28] (see also [5, 6]). These expressions were obtained by different methods using the properties of hypergeometric functions [15, 16, 19, 22], the hypervirial theorem [17, 18, 24, 27, 28], the hypervirial theorems and sum rules formulas [14] and using the second quantization method [25, 26]. It should be emphasized that explicit (non-recurrence) algebraic expressions for the vibrational dependence of matrix elements for an arbitrary function of internuclear distance are absent in the literature. Receipt of such expressions is actual problem for the development of algebraic methods of the study of vibrational dependences of various characteristics of diatomic molecules.

This article begins the cycle of papers devoted to the development of an algebraic approach to the determination of vibrational and vibrational−rotational dependences of different spectroscopic characteristics of the rovibrational states of diatomic molecules by means of the ladder operators. The purpose of this cycle is to: 1) develop the algebraic model for the vibrational matrix elements of quantum mechanical operators of fairly general form using the apparatus of the second quantization for diatomic molecules; 2) obtain missing in the literature algebraic expressions for the dependences of these matrix elements and different spectroscopic characteristics of diatomic molecules on vibrational and rotational quantum numbers of the combining states, and 3) make an algebraic analysis of the influence of adiabatic effects (anharmonicity, vibrational-rotational interaction) as well as non-adiabatic effects (electron-vibrational and electronic-rotational interactions) on these dependences of spectroscopic characteristics.

The present paper is devoted to the receiving algebraic expressions for matrix elements on arbitrary sufficiently differentiable function of internuclear distance $r$ on the vibrational wave functions of the potential curve whose explicit expressions for the creation and annihilation operators are known. The only condition whose fulfillment is essential for us is the fact that the action of creation and annihilation operators on the wave functions should give the wave functions of the Schrodinger equation with the same parameter values. Without going into details of the group theory description of these operators we still note that this situation occurs, in particularly, when these operators form the Weyl−Heisenberg algebra (case of the harmonic oscillator [5,29]) and the su(2) algebra (the Morse potential [20,30] and the modified Pöschl–Teller potential [21]). The purpose of the present paper is to obtain analytical formulas for the matrix elements $\langle v|f(r)|v'\rangle$ [1] for arbitrary values of vibrational quantum numbers of the combining states $v$ and $v'$ in terms of matrix elements on the wave functions of the ground vibrational state for these potentials as well as other possible potentials which satisfy to the conditions described above higher.

## 2. Notations

Consider non-relativistic Schrödinger equation for the vibrational wave function $\varphi_v(r)$ of the diatomic molecule

$$\hat{H}\varphi_v(r) = E_v\varphi_v(r) \ ,$$

$E_v$ is the energy of $v$−th vibrational level. Disregarding the rotation of the molecule as well as the non−adiabatic effects of intermolecular interactions [2] the Hamiltonian of the molecule is given by

$$\hat{H}(r) = -\frac{\hbar^2}{2\mu}\frac{d^2}{dr^2} + V(r),$$

where $V(r)$ is the potential energy curve of the electronic state under investigation; $\mu$ is the reduced mass of the molecule.

The factorization method is consists in to represent $V(r)$ in the form of an analytical potential of some parameters $\vec{\xi} = (\xi_1, \xi_2, ..., \xi_n)$

$$V(r) = U(\vec{\xi}, r) \tag{1}$$

and reduce the Schrodinger equation with this potential to the two first order differential equations [5, 6]. This is achieved by the substituting $r$ by a dynamic variable $y$ for which creation and annihilation operators can be constructed:

---

[1] For greater visibility here and further in the article the quantum numbers describing the electronic state of the molecule are not given since the subject of this study is namely vibrational dependences of matrix elements.

[2] The influence of vibration-rotation interactions and also non-adiabatic effects on vibrational and rotational dependences of matrix elements will be analyzed in our next article.

$$\hat{K}^+ \varphi_\nu(y) = \hat{a}_\nu^+ \varphi_{\nu+1}(y) \ , \tag{2}$$

$$\hat{K}^- \varphi_\nu(y) = \hat{a}_\nu^- \varphi_{\nu-1}(y) \ . \tag{3}$$

The action of the annihilation operator on the wave function of the ground vibrational state is given by:

$$\hat{K}^- \varphi_0(y) = 0 \ . \tag{4}$$

In general the creation and annihilation operators have the form (see [5]) [3]:

$$\hat{K}^\pm = \frac{d}{dy} \hat{b}_\nu^\pm + \hat{c}_\nu^\pm(y) \ . \tag{5}$$

Operator coefficients $\hat{a}_\nu^\pm$, $\hat{b}_\nu^\pm$ and functions $\hat{c}_\nu^\pm(y)$ depend in general on the vibrational quantum number of the wave function on which they act, as well as the parameters of the potential of the molecule. For example in the case of the Morse potential such parameters are the dissociation energy $D_e$ and the parameter $\alpha$ staying in the exponents of this potential (see Table). The result of the action these operators on the wave functions will be no longer denoted as operator but the numerical coefficients $a_\nu^\pm$, $b_\nu^\pm$ and functions $c_\nu^\pm(y)$. The explicit form of expressions for the coefficients $a_\nu^-$ and $b_\nu^-$ as well as the dependence of dynamic variable on internuclear distance $y=y(r)$ for the harmonic oscillator and the Morse potential necessary for further analysis are given in Table.

Now let us turn to an analysis of the matrix elements $\langle v | f(r) | v' \rangle$. In spectroscopic studies the function $f(r)$ can corresponds to dependences on the internuclear distance of the dipole moments (also higher order moments) of the electric (and magnetic) transitions, the dipole moments of molecule, the tensor coefficients of the electrical polarizability, the magnetic susceptibility and other physical characteristics of the molecule.

---

[3] In the literature these operators are sometimes given in the form $\hat{K}^\pm = g(x) \dfrac{d}{dx} \hat{b}_\nu^\pm + \hat{c}_\nu^\pm(x)$ (see [5]). The operators given in this form can be represented in the form described by formula (5) by means of the variable substitution $y = \displaystyle\int_{x_0}^{x} (1/g(z)) dz$ .

**Table.** The form of the dependence $y=y(r)$, the coefficients $a_v^-$, $b_v^-$ and the wave function of the ground vibrational state $\varphi_0(y)$ for the harmonic oscillator and the Morse potential (according to data from [5, 29]); here $q = (8\mu D_e)^{1/2}/\alpha\hbar$.

| Parameters | Potentials | |
|---|---|---|
| | Harmonic oscillator | Morse potential |
| $U(\vec{\xi}, r)$ | $\dfrac{\hbar^2}{2\mu}\alpha^4(r-r_e)^2$ | $D_e[\exp(-2\alpha(r-r_e)) - 2\exp(-\alpha(r-r_e))]$ |
| $y=y(r)$ | $\alpha(r-r_e)$ | $q\exp(-\alpha(r-r_e))$ |
| $a_v^-$ | $\sqrt{v}$ | $\sqrt{v(q-v)}$ |
| $b_v^-$ | $\dfrac{1}{\sqrt{2}}$ | $-(q-2v)\sqrt{\dfrac{q-2v+1}{q-2v-1}}$ |
| $\varphi_0(y)$ | $\dfrac{\alpha^{1/2}}{\pi^{1/4}}\exp(-y^2/2)$ | $\left(\dfrac{\alpha}{\Gamma(q-1)}\right)^{1/2} y^{(q-1)/2}\exp(-y/2)$ |

## 3. Diagram technique for determination of vibrational dependences of matrix elements

Our goal is to express the matrix elements $\langle v|f(y)|v'\rangle$ on the wave functions of states with arbitrary values of the vibrational quantum numbers $v$ and $v'$ in terms of matrix elements on the wave functions of the ground vibrational state. Without loss of generality consider the matrix elements $\langle v|f(y)|v+m\rangle$ where $m$ is an integer satisfying to the condition $-v \le m \le (v_{max} - v)$; $v_{max}$ is the maximum value of the vibrational quantum number for bound vibrational states of the potentials having the dissociation threshold, and $v_{max} = \infty$ for potentials are without such a threshold (for example, the harmonic oscillator). First, consider the case $m \ge 0$.

According to the definition of the commutator the following relation takes place

$$\hat{K}^- f(y) = \left[ \hat{K}^- f(y) \right] + f(y)\hat{K}^- \ , \qquad \qquad . \ (6)$$

where $\left[ \hat{A}\hat{B} \right]$ is the commutator of the operators $\hat{A}$ and $\hat{B}$. The expression for the commutator in the formula (6) has the form:

$$\left[ \hat{K}^- f(y) \right] = \left( \frac{df(y)}{dy} \right) \hat{b_v} \ .$$

Using this relation and Eqs. (2), (3) and (6) and the mutually conjugation of the operators $\hat{K}^+$ and $\hat{K}^-$ we obtain

$$\langle v|f(y)|v+m \rangle = \frac{1}{a_v^-} \ b_{v+m}^- \left\langle v-1 \left| f^{(1)}(y) \right| v+m \right\rangle + a_{v+m}^- \left\langle v-1 | f(y) | v+m-1 \right\rangle \ . \quad (7)$$

Here and further it is used the following notation

$$f^{(k)}(y) = \frac{d^k}{dy^k} f(y) \ .$$

Note that Eq. (7) contains numerical coefficients $a_v^-$, $a_{v+m}^-$ and $b_{v+m}^-$ instead corresponding operators.

It can be seen from Eq. (7) that the using creation and annihilation operators allows decreasing value of the vibrational quantum number $v$ on 1 for the wave function of one from the combining vibrational states. Applying formula (7) to the matrix elements to the right in this formula we decrease value of the vibrational quantum number of the wave function of the vibrational state again on 1. Thus, the sequential application $v$ times of this formula leads decreasing from $v$ to 0 value of the vibrational quantum number of the wave function of one from the combining vibrational states. This allows us to express the matrix elements $\langle v|f(y)|v+m \rangle$ as a linear sum of the matrix elements $\left\langle 0 \left| f^{(k)}(y) \right| \tilde{v} \right\rangle$ (where $\tilde{v} = m$, $m+1$, ..., $(m+v)$ and $k=0, 1, 2, ..., v$). Taking into account this we will be carried out the solution of our task in two stages. At first, we will obtain a formula expressing the matrix elements $\langle v|f(y)|v+m \rangle$ in terms of the matrix elements $\left\langle 0 \left| f^{(k)}(y) \right| \tilde{v} \right\rangle$. Then we will obtain a formula expressing the matrix elements $\left\langle 0 \left| f^{(k)}(y) \right| \tilde{v} \right\rangle$ in terms of the matrix elements $\left\langle 0 \left| f^{(k)}(y) \right| 0 \right\rangle$.

To obtain the required expression and also illustrate the effect of creation and annihilation operators we use the following diagram technique. We construct a diagram with points $(l, k)$ which conform to the matrix elements of the form $\left\langle l \left| f^{(k-l)}(y) \right| k+m \right\rangle$ (see Fig.). Value $k$ is plotted on the abscissa in descending order from $v$ to 0. Value $l$ is plotted on the ordinate in ascending order from 0 up to $v$. In accordance with Eq. (7) it is possible to go from each point $(l, k)$ to two points $(l-1, k)$ and $(l-1, k-1)$ that corresponds to the first and second terms on the right in Eq. (7). The vertical arrow ("down") corresponds to the transition from point $(l, k)$ to point $(l-1, k)$ and the diagonal arrow ("down and right") corresponds to the transition from point $(l, k)$ to point $(l-1, k-1)$ (see the supplementary figure at the top right corner of Fig.). With regard to Eq. (7) these transitions corresponds to the multiplication on the coefficients $b^-_{k+m}/a^-_l$ and $a^-_{k+m}/a^-_l$ correspondingly. Let us introduce the definition of a path vector as a continuous line connecting the points of the diagram by a sequence of vertical and diagonal arrows. Then the result of repeated application of formula (7) is a set of path vectors. It is need to transit from points $(v, v)$ to the line $l$=0 ($X$-axis). This corresponds to a set of path vectors limited by the area enclosed by the triangle ABC whose vertices are points $(v, v)$, $(0, v)$ and $(0, 0)$ correspondingly (see Fig.). Thus, the required expression implies the summarizing of matrix elements through all these path vectors.

This summarizing can be fulfilled as follows. Initially we fix point $(0, j)$ (point D) on the $X$-axis (Fig.). This point corresponds to the matrix element $\left\langle 0 \left| f^{(j)}(y) \right| j+m \right\rangle$ in the required expression. The set of all possible (in accordance with formula (7)) path vectors from point $(v, v)$ (point A) towards point $(0, j)$ (point D) is limited by a parallelogram whose vertices are points A and D and points $(v-j, v)$ and $(j, j)$ designated as points E and F accordingly. In Fig. this parallelogram AEDF is selected by the dot-dashed line. As an illustration it is shown one on Fig. a possible path vector $AG_1G_2\ldots G_{v-1}D$ connecting point $(v, v)$ and point $(0, j)$. It should be noted that as only the tip of the path vector falls on the line ED or FD then there is only one further route to get to the point D namely to go along one of these lines.

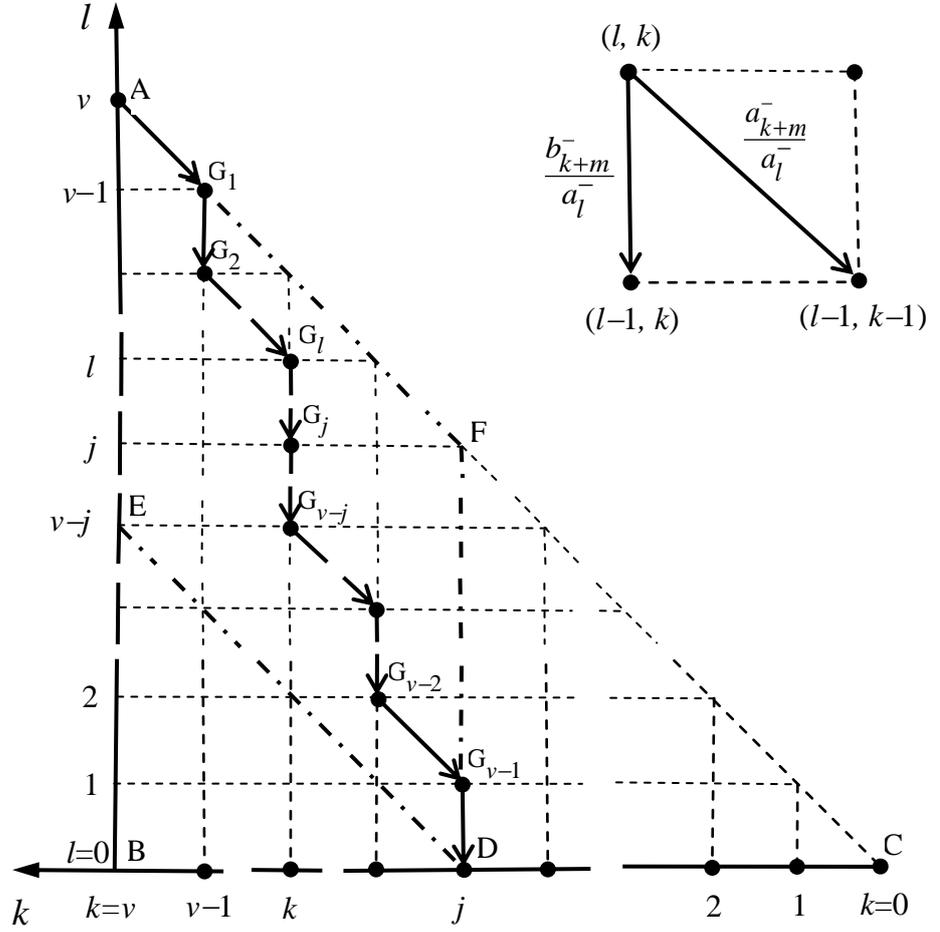

**Fig.** The diagram illustrating the sequential application of formula (7). Points $(l, k)$ correspond to matrix elements $\left\langle l \left| f^{(k-l)}(y) \right| k+m \right\rangle$. Value $k$ is plotted on the abscissa and value $l$ is plotted on the ordinate. The line $\mathrm{AG_1G_2...G_l...G_j...G_{v-j}...G_{v-2}G_{v-1}D}$ corresponds to one of the possible path vectors appearing at the transformation the matrix element $\left\langle v \left| f(y) \right| v+m \right\rangle$ into the matrix elements $\left\langle 0 \left| f^{(j)}(y) \right| j+m \right\rangle$. A diagram illustrating a single action of formula (7) is in the upper right corner of the figure. The vertical arrow from point $(l, k)$ to point $(l-1, k)$ corresponds to the first term on the right in brackets in formula (7) and the diagonal arrow from point $(l, k)$ to point $(l-1, k-1)$ corresponds to the second term on the right in brackets in this formula.

Using this notation let us analyze the coefficients staying before the matrix elements in formula (7) into the required expression.

(i) Any vector path that lies within parallelogram AEDF contains $v$ arrows in all that corresponds to the application $v$ times of formula (7). Therefore consideration the factors before brackets in formula (7) gives the factor $(A(v))^{-1}$ in the required expression where

$$A(v) = \prod_{k=1}^{v} a_k^{-} \ . \tag{8}$$

(ii) Since any path vector that lies within the AEDF area contains $v-j$ diagonal arrows taking into account of the right term in brackets in formula (7) gives us the factor (for $0 \le j \le v-1$):

$$A_j(v,m) = \prod_{k=j+1}^{v} a_{k+m}^{-} \ . \tag{9}$$

For $j = v$ the coefficient $A_v(v,m) = 1$ since the path vector contains only vertical arrows in this case (see Fig.).

(iii) Take into account the left term in brackets in formula (7) which is shown arrows "down" on Fig. Any path vector that lies within the AEDF area contains $j$ vertical arrows. Note that the first arrow "down" on a vector path that lies in the AEDF area can have any abscissa value $i_1$ from $j$ to $v$. The second arrow "down" on a path vector should not be to the left of the previous arrow "down" on this path vector. Therefore this second arrow "down" can have any the abscissa value $i_2$ from $j$ up to $i_1$. Continuing this reasoning further allows us to draw conclusion that the last $j$-th arrow "down" on the path vector can have any abscissa value $i_j$ from $j$ up to $i_{j-1}$. Thus, considering coefficients $b_k^{-}$ at the application $v$ times of the formula (7) gives the factor

$$B_j(v,m) = \left( \sum_{i_1=j}^{v} b_{i_1+m}^{-} \sum_{i_2=j}^{i_1} b_{i_2+m}^{-} \sum_{i_3=j}^{i_2} b_{i_3+m}^{-} \cdots \sum_{i_j=j}^{i_{j-1}} b_{i_j+m}^{-} \right) . \tag{10}$$

The joint taking into account of the results (i)–(iii) allows us to obtain the coefficient before the matrix element $\left\langle 0 \left| f^{(j)}(y) \right| j+m \right\rangle$ in the required expression. This coefficient is equal to the factor $A_j(v,m) B_j(v,m) \big/ A(v)$ .

To obtain an expression for the matrix element $\langle v | f(y) | v+m \rangle$ it is only need to sum over $j$ from 0 up to $v$. At this summarizing point D in Fig. should run all the points lying on the $X$-axis from point C towards point B. Then we obtain the requared expression:

$$\langle v | f(y) | v+m \rangle = \frac{1}{A(v)} \sum_{j=0}^{v} A_j(v,m) B_j(v,m) \left\langle 0 \left| f^{(j)}(y) \right| j+m \right\rangle . \qquad (11)$$

At last we express the matrix elements to the right in Eq. (11) through the matrix elements of the wave functions of the ground vibrational level. Using the Eqs. (2)–(6) and mutually conjugation of the operators $\hat{K}^+$ and $\hat{K}^-$ we obtain

$$\left\langle 0 \left| f^{(j)}(y) \right| j+m \right\rangle = \frac{(b_0^-)^{j+m}}{A(j+m)} \left\langle 0 \left| f^{(2j+m)}(y) \right| 0 \right\rangle . \qquad (12)$$

Expressing the rightmost term in Eq. (11) with regard to formula (12) we finally obtain

$$\langle v | f(r(y)) | v+m \rangle = \sum_{j=0}^{v} \gamma_j(v,m) \left\langle 0 \left| f^{(2j+m)}(y) \right| 0 \right\rangle , \qquad (13)$$

here

$$\gamma_j(v,m) = (b_0^-)^{j+m} \frac{A_j(v,m) B_j(v,m)}{A(j+m) A(v)} . \qquad (14)$$

So far we considered the case $m \geq 0$. If $-v \leq m < 0$ it is need to replace $v$ by $v+m$ and $m$ by $|m|$ in the right-hand side of Eqs. (13) and (14). Denote $v_m = \min(v, v+m)$. Then for the general case of arbitrary values $m$ satisfying the condition $0 \leq v+m \leq v_{max}$ Eq. (13) can be written as:

$$\langle v | f(r) | v+m \rangle = \sum_{j=0}^{v_m} \gamma_j(v_m, |m|) \left\langle 0 \left| f^{2j+|m|}(y) \right| 0 \right\rangle . \qquad (15)$$

Eqs. (14) and (15) are the main result of the present paper. The expressions for particular cases of matrix elements can be derived as a consequence of these formulas. For example the formula for the diagonal matrix elements ($v' = v$) corresponding to the expectation of a certain physical characteristics (described by the function $f(r)$) in the state with vibrational quantum number $v$ is obtained from Eqs. (14) and (15) by substituting $m=0$ and $v_m = v$.

## 4. Analysis

It can be seen that the expression for the matrix elements $\left\langle v \left| f(r) \right| v+m \right\rangle$ (see Eq. (15)) is defined (see Eqs. (8)–(10)) by: 1) the coefficients $a_i^-$ ($i=1, 2, \ldots, v_m + |m|$) and $b_i^-$ ($i=|m|, |m|+1, \ldots, v_m + |m|$) which, in turn, are determined by the explicit form of the annihilation operator (see Eqs. (3) and (4)) and the parameters of the potential curve ($\xi_1, \xi_2, \ldots, \xi_n$) (Eq. (1)); 2) the wave function of the ground vibrational state, 3) the $p$-th order derivatives of function $f(y)$ ($p=|v-v'|, |v-v'|+1, \ldots, (v+v')$) (see Eq. (15)) that are determined by the analytic properties of this function. It is important to note that to calculate of matrix elements $\left\langle v \left| f(r) \right| v' \right\rangle$ for arbitrary values $v$ and $v'$ it is enough to have information about analytic properties of the function $f(y)$ only for those value of $r$ where the wave function of the ground vibrational state is not negligible.

All necessary data for algebraic calculations of various characteristics of the vibrational states of diatomic molecules by Eq. (15) for the harmonic oscillator and the Morse potential are given in Table. These data were obtained using the data of [5, 29]. The coefficients $\gamma_j(v_m, |m|)$ can be easily calculated and tabulated for a range of the molecule parameters ($\xi_1, \xi_2, \ldots, \xi_n$) of interest in spectroscopic studies if the explicit form of the operators $\hat{K}^+$ and $\hat{K}^-$ is known.

For further analysis expand function $f(y)$ in a Taylor series in the neighborhood of a point $y_0$ (for example it can be the value $y$ corresponding to equilibrium internuclear distance $r_e$):

$$f(y) = \sum_{k=0}^{\infty} \frac{f^k(y_0)}{k!} (y - y_0)^k \ . \tag{16}$$

Take into account that

$$f^{2j+|m|}(y) = \sum_{k=0}^{\infty} \frac{f^{k+2j+|m|}(y_0)}{k!}(y-y_0)^k .$$

Substituting this expression into Eq. (15) we obtain

$$\langle v|f(y)|v+m\rangle = \sum_{k=0}^{\infty} \zeta_k(v_m,|m|)\langle 0|(y-y_0)^k|0\rangle , \qquad (17)$$

here

$$\zeta_k(v_m,|m|) = \frac{1}{k!}\sum_{j=0}^{v_m}\gamma_j(v_m,|m|)f^{(k+2j+|m|)}(y_0) . \qquad (18)$$

It can seen that to calculate the matrix elements it is enough to have information about: 1) the coefficients $a_i^-$ and $b_i^-$ (including in factors $\gamma_j(v_m,|m|)$ (see Eq. (14)); 2) the matrix elements only on the wave functions of the ground vibrational state (see Eq. (17)); 3) the $p$–order derivatives of function $f(y)$ in a point $y_0$ (see Eq. (18)), where $p \geq |m|$. Matrix elements $\langle 0|(y-y_0)^k|0\rangle$ can be determined analytically for the special cases of potentials. The values of the derivatives $f^{(p)}(y_0)$ occurring in Eq. (15) can be found semi-empirically or using the results of *ab initio* calculations. It is important to note that each differentiation of the function is associated with appearance of some scale factor that depends on molecule parameters. This leads to a certain hierarchy of the factors $\zeta_k(v_m,|m|)$. Consideration of this hierarchy allows limiting the number of terms in Eq. (17) by several terms.

As an example of efficiency of obtained algebraic expressions consider one-dimensional harmonic oscillator and obtain a formula for the vibrational dependences of matrix elements for the case. Using Eqs. (8)–(10), (14) and the data given in Table we obtain:

$$\gamma_j^{|m|}(v) = \left(\frac{1}{\sqrt{2}\alpha}\right)^{2j+|m|}\frac{\sqrt{(v_m+|m|)!}}{(j+|m|)!\sqrt{v_m!}}\left(\sum_{i_1=j}^{v_m}\sum_{i_2=j}^{i_1}\sum_{i_3=j}^{i_2}\ldots\ldots\sum_{i_j=j}^{i_{j-1}}1\right) . \qquad (19)$$

It is not difficult to show that the number of terms defined by the $j$-sums in Eq. (19) equals the binomial coefficient $C_{v_m}^j$. This corresponds to the number of all possible different paths vector from point A towards point D on Fig. Thus, we obtain the required formula

$$\langle v|f(r)|v+m\rangle = \sqrt{v_m!(v_m+|m|)!} \sum_{j=0}^{v_m} \frac{1}{2^{j+\frac{|m|}{2}}\alpha^{2j+|m|}j!(j+|m|)!(v_m-j)!} \langle 0|f^{(2j+|m|)}(r)|0\rangle.$$

We get another form of this expression expanding $f(r)$ in a Taylor series with regard to Eq. (16). Taking into account that the function $\varphi_0(y)$ of the harmonic oscillator is even (see Table) and replacing the integration over the interval $0 \le r \le \infty$ on the integration on the interval $-\infty \le r \le \infty$ (since the difference between these integrals is negligible for the real potentials of molecules) the following formula is finally obtained

$$\langle v|f(r)|v+m\rangle = \sqrt{v_m!(v_m+|m|)!} \sum_{k=0}^{\infty} \sum_{j=0}^{v_m} \frac{f^{\left(2k+2j+|m|\right)}(r_e)}{2^{2k+j+\frac{|m|}{2}}\alpha^{2k+2j+|m|}(j+|m|)!(v_m-j)!j!k!} \quad . \text{(20)}$$

In this it is used the following relation for the harmonic oscillator

$$\int_{-\infty}^{\infty} (\varphi_0(y))^2 y^{2k} dy = \frac{1}{\sqrt{\pi}} \Gamma\left(k+\frac{1}{2}\right) = \frac{(2k)!}{2^{2k}k!} \quad ,$$

here $\Gamma(z)$ is the Gamma function.

It should be noted that Eq. (20) for the case $\alpha = 1$, $r_e = 0$ and $m \ge 0$ was obtained earlier by using hypervirial theorem and a technique of differentiation with respect to the parameter [12] and the binomial formula for operators with Cauchy's integral theorem and Baker-Campbell-Hausdorff theorem [13]. It should be noted that there is a misprint in formula (4.3) from [12]: in this formula the $f^{(2j+n-m-2r)}(0)$ must be replaced by the $f^{(2j+n-m+2r)}(0)$. This misprint has been found in [13].

A wide range of physical and chemical problems (molecular plasma physics, plasma chemistry, etc.) requires information about the total set of matrix elements for all possible values of vibrational quantum numbers of the combining states. An important consequence of the expressions obtained in present paper is the possibility obtaining these date calculating only $(2v_{\max}+1)$ matrix elements (see Eq. (15))[4] instead the computation $(v_{\max}+1)\times(v_{\max}+2)/2$ independent values of matrix elements taking place at numerically salvation of the Schrödinger equation. Thus, using the algebraic expressions obtained in present paper can reduce significantly (approximately $v_{\max}/4$ times) the number of necessary calculations of matrix elements compared to the number of calculations using numerical methods. Even for comparatively light diatomic molecules (except the hydrogen molecule and hydrides) using the obtained expressions allows reduce the number of necessary calculations more than one order of magnitude.

## 5. Conclusion

The diagram technique describing the actions of the annihilation operator at the consideration of vibrational dependences of matrix elements has been developed. The algebraic formulas for the matrix elements with arbitrary values of the vibrational quantum numbers of the combining states for arbitrary sufficiently differentiable functions of the internuclear distance are obtained in terms of matrix elements on the wave functions of the ground vibrational state. These formulas for the case of anharmonic potential curves were previously absent in the literature known to us. It is important to emphasize that proposed method for determining of vibrational dependences of matrix elements is a direct, simple and intuitive unlike the techniques developed earlier and does not require using any special theories of quantum mechanics (hypervirial, Hellmann-Feynman and others) and hypergeometric functions. The only required information reduces to knowing the explicit form of the annihilation operator and the wave function of the ground vibrational state of the molecule. All necessary data for such algebraic calculations of the vibrational dependencies of matrix elements for the harmonic oscillator and the Morse potential are given. The efficiency of the diagram technique method is illustrated by the example of matrix elements for the harmonic oscillator.

Obtained formulas can be used to determine the absolute values and the vibrational dependence of various radiative, electrical, magnetic and other characteristics of the vibrational states of diatomic molecules as well as at the study of various spectroscopic characteristics of polyatomic molecules, for example, using the local mode model [31]. These formulas give the opportunity for obtaining analytical expressions for the vibrational dependence of matrix elements on the functions that have more complex form than those studied previously, for example, the functions having the form of Padé approximation. Functions of this type are widely used to describe the dipole moments of the hydrogen halides and other diatomic molecules (see [32] and references therein). It should be noted that obtained formulas can be used to analyze the vibrational dependences of various characteristics not only for the ground electronic states but also the electronically excited states of molecules (including the branching ratios of spontaneous emission, radiative lifetimes, Λ-doubling, *g*-factors, etc.) for example with the help of the sum rule formulas [33].

---

[4] This result has been pointed out by Prof. V. V. Smirnov.

## Aknowlegement

The author is grateful to Prof. V. V. Smirnov for his valuable comments and observations (see footnote 4).